\documentclass[12pt]{article}
\usepackage{epsf}

\def\pin{$\mbox{}$\indent}  

\newcounter{shimeqsno} 

\def\bsigma{\mbox{\boldmath $\sigma$}}
\def\bmu{\mbox{\boldmath $\mu$}}

\def\bxi{\mbox{\boldmath $\xi$}}
\def\bpsi{\mbox{\boldmath $\psi$}}
\def\bgamma{\mbox{\boldmath $\gamma$}}
\def\bneta{\mbox{\boldmath $\eta$}}

\def\bm{{\bf m}}

\def\bk{{\bf k}}
\def\bs{{\bf s}}

\def\bR{{\bf R}}

\def\sign{\mbox{\rm sign}}

\begin{document}
\setcounter{page}{0}
\begin{titlepage} 
\title{Statics and dynamics of an Ashkin-Teller neural network with low
loading}
\author{ D.~Boll\'e 
     \thanks{ e-mail:
        desire.bolle@fys.kuleuven.ac.be \hfill \break  \indent  $^{(1)}$
        Also at Interdisciplinair Centrum voor Neurale
         Netwerken, K.U.Leuven} $^ {(1)}$
     and P.~Koz{\l}owski 
     \thanks{ permanent address: Computational Physics
       Division, Institute of Physics, A. Mickiewicz University,
        ul. Umultowska 85, PL 61-624 Pozna\'{n}, Poland \hfill \break
      e-mail: piotr@tfdec1.fys.kuleuven.ac.be} $^{(1)}$
                 \\ \\
Instituut voor Theoretische Fysica,
K.U. Leuven \\
B-3001 Leuven, Belgium\\}
\date{}
\maketitle
\thispagestyle{empty}

\begin{abstract}
\noindent
An Ashkin-Teller neural network, allowing for two types of neurons is
considered in the case of low loading as a function of the strength of the
respective couplings between these neurons. The storage and retrieval of
embedded patterns built from the two types of neurons, with
different degrees of (in)dependence is studied.
In particular, thermodynamic properties including the existence and
stability of Mattis states are discussed. Furthermore, the dynamic
behaviour is examined by deriving flow equations for the macroscopic
overlap. It is found that for {\it linked} patterns the model
shows better retrieval properties than a corresponding Hopfield model.
\end{abstract}
\vspace*{1cm}
\end{titlepage}

\section{Introduction}
\label{sec:intro}
\pin

One of the best known physical models for neural networks is the Hopfield
model \cite{hop}. In theoretical investigations of network properties,
e.g., the retrieval of learned patterns, it plays a similar role as the
Ising model does in the theory of magnetism.
Extensions of this model to multi-state neurons have received a lot of
attention recently (see, e.g., \cite{GK} - \cite{BRS}
and the references cited therein). Thereby the ability to store and
retrieve so-called grey-toned and coloured patterns has been investigated.

In this work we consider another extension of the Hopfield model to
allow for multi-functional neurons. 
The specific model we have in mind is the neural network
version of the Ashkin-Teller spin-glass (\cite{chri}-\cite{nob}). Indeed,
on the one hand the Ashkin-Teller model has two different kinds of neurons
(spins) at each site interacting with each other. This allows us to
interprete this model as a neural network with two types of neurons having
different functions. On the other hand, this Ashkin-Teller neural network
(ATNN) can be considered as a model consisting out of two interacting
Hopfield models.

We expect the behaviour of the ATNN to be different from the one of the
Hopfield model in a non trivial way. One of the things we want to find out,
e.g., is whether this (four-neuron) interaction between the two types of
neurons can improve the retrieval process for embedded patterns
built from these two types of neurons. We will see, indeed, that for a
particular choice of this interaction term the retrieval quality of the
embedded patterns is very high in comparison with a corresponding
Hopfield model. 
Therefore, independent of the possible biological relevance of this
model, if any, such a study is interesting from the pure physical point 
of view.

In this work we consider both the thermodynamic and dynamic properties
of this model in the case of loading of a finite number of patterns.

The rest of this paper is organized as follows. In section \ref{sec:mod}
the ATNN model is introduced. Section \ref{sec:met} discusses the methods
used for analyzing both the equilibrium properties and the dynamics of the
model. In particular, fixed-point equations as well as flow equations for
the relevant macroscopic overlap order parameters are derived. In section
\ref{sec:res} numerical
solutions of these equations are discussed for a representative set of
network parameters. The retrieval properties of embedded
patterns with different degrees of dependencies are compared.
Section \ref{sec:con} presents the main conclusions.

\section{The model}
\label{sec:mod}
\pin

We consider a network of $N$ sites. At each site we have two different
types of binary neurons, $s_i$ and $\sigma_i \,\, , i=1,\ldots,N$. The two
types of neurons interact via a four-neuron term $s_i s_j \sigma_i
\sigma_j$. The infinite-range hamiltonian reads
\begin{equation}
   H=-\frac{1}{2} \sum_{i,j} [J_{ij}^{(1)} s_i s_j +
     J_{ij}^{(2)} \sigma_i \sigma_j + J_{ij}^{(3)} 
                               s_i s_j \sigma_i \sigma_j]
         \label{eq:atnn0}
\end{equation}

In this network we want to store a finite number of patterns, $p$, also of
two different types, i.e., $\bxi_i=\{\xi_i^\mu\}, \mu=1,\ldots,p$ and
$\bneta_i=\{\eta_i^\nu\}, \nu=1,\ldots,p$, which are supposed to be
independent identically distributed random variables (i.i.d.r.v.)
taking the values $+1$ or $-1$ with probability $1/2$. To build in the
capacity for learning and retrieval in this network its stable
configurations must be correlated with the configurations determined by the
learning process. This can be accomplished by taking the Hebb learning rule
for the interactions
\begin{equation}
     J_{ij}^{(1)}= \frac{1}{N} J_1 \sum_{\mu=1}^p \xi_i^\mu \xi_j^\mu,
      \quad
     J_{ij}^{(2)}= \frac{1}{N} J_1 \sum_{\mu=1}^p \eta_i^\mu \eta_j^\mu,
     \quad
 J_{ij}^{(3)}= \frac{1}{N} J_3 \sum_{\mu=1}^p \gamma_i^\mu \gamma_j^\mu \,,
     \label{eq:hebb}
\end{equation}
where the $\bgamma_i=\{\gamma_i^\mu\}, \mu=1, \ldots, p$ are also
i.i.d.r.v. taking the values $+1$ or $-1$ with probability $1/2$.

At this point some remarks are in order. Firstly, we have taken the
strength of the two types of patterns to be equal, meaning that the
ATNN model is isotropic. Secondly, it is clear that the behaviour of this
model (\ref{eq:atnn0})-(\ref{eq:hebb}) might depend on the fact whether
the $\bgamma$ are taken to be independent from the $\bxi$ and the
$\bneta$ or not. The following cases will be distinguished:
\begin{enumerate}
\item unlinked patterns
\begin{enumerate}
\item $\bxi_i, \bneta_i$ and $\bgamma_i$ are i.i.d.r.v.
\item $\bxi_i= \bneta_i$ and $\bgamma_i$ are i.i.d.r.v.
\item $\bxi_i= \bneta_i=\bgamma_i$ is i.i.d.r.v.
\end{enumerate}
\item linked patterns \newline
$\gamma^{\mu}_i=\xi^{\mu}_i\eta^{\mu}_i$ with $\bxi_i$ and $\bneta_i$
i.i.d.r.v.
\end{enumerate}
We note that this ATNN model can also be considered as an assembly of two
single Hopfield models (when $J_3=0$), one in the $s_i$-neurons and one in
the $\sigma_i$-neurons interconnected via a four-neuron interaction (when
$J_3 \neq 0$). The study of coupled Hopfield networks has aroused some
interest in the literature before (e.g., \cite{VM}).

In the following we discuss both the thermodynamics and the dynamics of
this ATNN neural network with low loading.

\section{The method}
\label{sec:met}

\subsection{Statics}
\pin
Starting from the hamiltonian (\ref{eq:atnn0})-(\ref{eq:hebb}) and applying
standard techniques (linearization and the saddle-point method
\cite{prov}-\cite{ags}) the ensemble-averaged free energy is given by
\begin{eqnarray}
  &&f=\frac{1}{2}(J_{1}\bm_{1}^{2}+ J_{1}\bm_{2}^{2} +  J_{3}\bm_{3}^{2})
      \nonumber\\
   && -\frac{1}{\beta}\left\langle\!\left\langle
        \ln[4\cosh \beta L_1 \cosh \beta L_2 \cosh \beta L_3
               (1+ \tanh \beta L_1\tanh \beta L_2\tanh \beta L_3)]
               \right\rangle\!\right\rangle
      \nonumber\\
        \label{eq:free}
\end{eqnarray}
with
\begin{eqnarray}
    L_1= J_1\bxi\cdot\bm_1, \quad
    L_2= J_1\bneta\cdot\bm_2, \quad
    L_3= J_3\bgamma\cdot\bm_3 \,.
   \label{eq:els}
\end{eqnarray}
In the above the double brackets $\langle\!\langle \cdot \rangle\!\rangle$
denote the average over the distribution of the embedded patterns.
The $\bm_{\alpha}=\{m_{\alpha}^\mu\}, \mu =1, \ldots, p; \alpha=1,2,3$ are,
as usual, overlap order parameters defined by
\begin{equation}
  \bm_1=\frac{1}{N}\sum_{i=1}^N\bxi_i s_i, \quad
  \bm_2=\frac{1}{N}\sum_{i=1}^N\bneta_i \sigma_i, \quad
  \bm_3=\frac{1}{N}\sum_{i=1}^N\bgamma_i s_i\sigma_i \,.
  \label{eq:order}
\end{equation}
Here we remark that in the thermodynamic limit $N \rightarrow \infty$
and for finite loading $\alpha=0$ the diagonal terms in the couplings,
$J_{ii}$, do not play any role in the Hamiltonian (\ref{eq:atnn0}).

In fact our model can be considered as a special case of the general
spin-glass model presented in ref.~\cite{mor1} such that the expressions
(\ref{eq:free})-(\ref{eq:els}) can also be read off from there.

The fixed-point equations for the order parameters read
\begin{eqnarray}
\bm_{\alpha}=\left\langle\!\left\langle
      \frac{\bpsi_{\alpha}(\tanh \beta L_{\alpha}
      + \tanh \beta L_{\nu}\tanh \beta L_{\rho})}
      {1+\tanh \beta L_{\alpha}\tanh \beta L_{\nu}\tanh \beta L_{\rho}}
        \right\rangle\!\right\rangle\,,
        \label{eq:ec2}
\end{eqnarray}
where $\alpha, \nu, \rho  = 1,2,3$ are taken to be different and where
$\bpsi_{\alpha}$ is the embedded pattern corresponding to $\bm_{\alpha}$.

Since the study of this ATNN model is very involved we have restricted
ourselves here to a detailed treatment of the Mattis states (\cite{ags})
which are especially important from a neural network point of view. In our
case they are defined as those solutions of the fixed-point equations for
which not more than one component of each order parameter is different
from zero, e.g. $\bm_1=m_1(1,0,...,0)$, $\bm_2=m_2(1,0,...,0)$,
$\bm_3=m_3(1,0,...,0)$. These states are denoted by $m_1m_2m_3$ in the
sequel. We will see that those states are the only ones
which contribute to the thermodynamics of the system. Solutions with more
than one component being non-zero, i.e., mixture states will be
important for the dynamics if they are local minima of the free energy.

For zero temperature we note that the eqs.~(\ref{eq:ec2}) can be simplified
by replacing the $\tanh \beta L_{\alpha}$ by $\sign L_{\alpha}$. Then it
is straightforward to show that for each order parameter
\begin{equation}
              (\bm_{\alpha})^2 \leq 1 \,, \quad \alpha=1,2,3 \,,
\end{equation}
with the equality being satisfied for a one-component $\bm_{\alpha}$, and
that the ground-state energy is given by
\begin{equation}
               E= - \sum_{\alpha=1}^3 \frac12 \bm_{\alpha}^2 \,.
\end{equation}
In section \ref{sec:res} we report on the existence and stability of these
Mattis states as a function of the temperature.

\subsection{Dynamics}
\pin
The analysis outlined above enables us to find the local minima of the free
energy. But in order to find out how an arbitrary initial state of the
network changes in time and to what extent, if at all, one of the learned
patterns is approached, we derive a flow equation for the overlap
order parameters.

We consider sequential updating of the spins consistent with the detailed
balance condition. Hence we choose the following transition probabilities
for a spin flip at a certain time step
\begin{eqnarray}
  \omega_1 (s_i,\sigma_i)
           &\equiv&\omega (s_i\rightarrow -s_i)
         =\frac{1}{12}[1-\tanh (\beta (s_i h^{(1)}_i+
	                             s_i\sigma_i h^{(3)}_i))]
         \nonumber \\
  \omega_2 (s_i,\sigma_i)
           &\equiv&\omega (\sigma_i\rightarrow -\sigma_i)
         =\frac{1}{12} [1-\tanh (\beta (\sigma_i h^{(2)}_i+
	                              s_i\sigma_i h^{(3)}_i))]
          \nonumber \\
  \omega_3 (s_i,\sigma_i)
        &\equiv&\omega (s_i\rightarrow -s_i,\sigma_i\rightarrow - \sigma_i)
          =\frac{1}{12}[1- \tanh (\beta (s_i h^{(1)}_i + 
	  \sigma_i h^{(2)}_i))]
          \label{eq:trans}
\end{eqnarray}
with the $h_i$ appropriate local fields acting on the neurons in the
following way
\begin{eqnarray}
h^{(1)}_i=\frac{J_1}{N}\sum_j \sum_{\mu} \xi_i^{\mu}\xi_j^{\mu}s_j,\,\,
h^{(2)}_i=\frac{J_2}{N}\sum_j \sum_{\mu}
           \eta_i^{\mu}\eta_j^{\mu}\sigma_j,\,\,
h^{(3)}_i=\frac{J_3}{N}\sum_j \sum_{\mu} \gamma_i^{\mu}\gamma_j^{\mu}
         s_j \sigma_j \,,
\end{eqnarray}
where, as in the treatment of the statics we take $J_1=J_2$. At this
point we remark that in the thermodynamic limit $N \rightarrow \infty$
the diagonal terms in the couplings, $J_{ii}$, do not survive. Furthermore 
we note that the method used here is also valid in the case of unequal $J$.

We then consider the probability $p(\bs ,\bsigma;t )$ that the system is in
a state $\bs =(s_i,...,s_N)$, $\bsigma =(\sigma_1,...,\sigma_N)$ at time
$t$. It satisfies the master equation
\begin{eqnarray}
   \frac{\partial p(\bs,\bsigma;t )}{\partial t}
    &=&\sum_{i=1}^N \left \{
    \omega_1 (F_i^1s_i,\sigma_i)p(F_i^1\{\bs,\bsigma\};t)
      + \omega_2 (s_i,F_i^2\sigma_i) p(F_i^2\{\bs, \bsigma\};t)
         \right.
   \nonumber \\
    &+& \left. \omega_3 (F_i^3s_i,F_i^3\sigma_i) p(F_i^3\{\bs,\bsigma\};t)
                         \right.
   \nonumber \\
    &-& \left. p(\bs,\bsigma;t) [\omega_1(s_i,\sigma_i)
                    +\omega_2 (s_i,\sigma_i)+\omega_3 (s_i,\sigma_i)]
             \rule{0cm}{0.5cm} \right \} \,.
   \label{eq:master}
\end{eqnarray}
The operator $F_i^{\nu}$, $\nu =1,2,3$ acting on a configuration
$\{\bs ,\bsigma\}$ changes the sign of the following spins: $s_i$ for
$\nu =1$, $\sigma_i$ for $\nu =2$ and simultaneously $s_i$ and $\sigma_i$
for $\nu =3$.

>From this we want to derive a flow equation for the overlap order
parameters. Because of the multi-state character of the model
(due to the four-spin interaction term)
the summation over $i$ has to be carried out by generalizing the method of
submagnetizations or suboverlaps connected with partitions of the network
with respect to the built-in patterns \cite{bm}-\cite{hghk}. We note that
for Hopfield networks we do not need such a partitioning \cite{cool}.

First we introduce the following division of the network indices
\begin{equation}
  \{i\leq N\}=\bigcup_{\bk} I_{\bk}
              \quad I_{\bk}=\{i\leq N;\bk=\bk_i\}
              \quad \bk_i=(\bxi_i,\bneta_i,\bgamma_i)\,.
    \label{eq:dik}
\end{equation}
Then we define the so called submagnetisations or suboverlaps
\begin{equation}
   \mu_{\alpha ,\bk}(\bs, \bsigma)
    =\frac{1}{|I_{\bk}|}\sum_{i\in I_{\bk}}S_{\alpha,i}
       \quad \alpha =1,2,3  \quad
     S_{1,i}=s_i,~S_{2,i}=\sigma_i,~S_{3,i}=s_i\sigma_i~,
\end{equation}
which enables us to write the overlaps in the form
\begin{equation}
    m_{\alpha}^{\nu}=\sum_{\bk}\frac{|I_{\bk}|}{N}\mu_{\alpha ,\bk}
                k^{(\alpha-1)p+\nu}
        \quad \alpha =1,2,3, \quad \nu =1,...,p \,,
        \label{eq:qmu}
\end{equation}
where $|I_k|$ stands for the number of indices in the set $I_k$.
The number of vectors $\bk$ is equal to $2^{ap}$, which is much smaller
than the number $N$ of sites $i$ (when $N \rightarrow \infty$). 
The coefficient $a=3,2,1$ and $2$ for patterns of type $1(a), 1(b),
1(c)$ and $2$ respectively. So, an embedded random
pattern configuration can be assumed to be equally distributed over the
sets $I_k$ such that $|I_k|=2^{-ap}N + {\cal O}(N^{1/2})$.

Next we write down the probability $P(\bmu;t)$ that the
system is in a macroscopic state described by a set of submagnetisations
$\bmu(\bs ,\bsigma) \equiv \{\mu_{\alpha ,\bk}\}$
\begin{equation}
        P(\bmu;t)= \sum_{\{\bs ,\bsigma\}} p(\bs,\bsigma;t )
            \delta (\bmu -\bmu (\bs ,\bsigma)) \,.
\end{equation}
Then we arrive at the following master equation
(cfr.~eq.~(\ref{eq:master}))
\begin{equation}
  \frac{\partial P(\bmu;t)}{\partial t}
     =\sum_{\{\bs ,\bsigma\}}\sum_i\sum_{\nu}
     \omega_{\nu}(s_i,\sigma_i)p_t(\bs ,\bsigma )
     \left[\delta (\bmu -\bmu (F_i^{\nu}\{\bs ,\bsigma \}))-
            \delta (\bmu -\bmu (\bs ,\bsigma ))\right] \,.
            \label{eq:pmu}
\end{equation}
The action of the operator $F_i^{\nu}$ can be specified further by writing
\begin{equation}
  \bmu (F_i^{\nu}\{\bs ,\bsigma \})
       =\bmu (\bs ,\bsigma )- \bR^{\nu}(s_i,\sigma_i)\,.
\end{equation}
It is then straightforward to check that certain elements of the matrix
$[\bR^{\nu}(s_i,\sigma_i)]_{\alpha,\bk}$ are zero, viz.
\begin{equation}
   R^1_{2,\bk}=R^2_{1,\bk}=R^3_{3,\bk}=0 \quad \forall \bk, i\leq N \,.
   \label{eq:R0}
\end{equation}
Furthermore, for $\nu$ and $\alpha$ different from these specific values
in (\ref{eq:R0}) we have
\begin{eqnarray}
  && R^{\nu}_{\alpha ,\bk}=\frac{2}{|I_{\bk }|}S_{\alpha,i}
            \quad  \mbox{if}\,\,\, i \in I_{\bk}\\
  && R^{\nu}_{\alpha ,\bk}=0 \quad \mbox{if}\,\,\, i \not \in I_{\bk} \,.
\end{eqnarray}
In the thermodynamic limit $N \rightarrow \infty$ the parameters $\bmu$
become continuous variables. Following \cite{bm} we then write for an
arbitrary smooth function $\phi(\bmu)$
\begin{eqnarray}
  \frac{\partial \langle \phi(t)\rangle}{\partial t}
  && \equiv
  \int d \bmu \, \phi(\bmu) \frac{\partial P(\bmu;t )}{\partial t}
  \nonumber \\
  && = \sum_{\{\bs ,\bsigma\}} \sum_i\sum_{\nu}
  \int d\bmu^{'}
    p(\bs,\bsigma;t ) \delta (\bmu^{'} -\bmu (\bs ,\bsigma))
    \omega_{\nu}(s_i,\sigma_i)
    \nonumber\\
  && \times \left[\phi (\bmu^{'}-\bR^{\nu}(s_i,\sigma_i))
                                    - \phi (\bmu^{'})\right] \,.
\end{eqnarray}
Making an expansion of $\phi$  around
$\bmu^{'}$ and doing a partial integration with respect to
$\bmu^{'}$ we arrive at
\begin{eqnarray}
 \frac{\partial \langle \phi(t)\rangle}{\partial t}
 = \int d\bmu^{'}\, \phi (\bmu^{'})
   \sum_{\alpha,\bk}\frac{\partial}{\partial \mu^{'}_{\alpha ,\bk}}
   \left( \sum_{\{\bs ,\bsigma\}} \sum_i\sum_{\nu}
     p(\bs,\bsigma;t ) \delta (\bmu^{'} -\bmu (\bs ,\bsigma))
        \right.   \nonumber \\
    \left. \times  R_{\alpha ,\bk }^{\nu}(s_i,\sigma_i) \,
     \omega_{\nu}(s_i,\sigma_i) \rule{0cm}{0.6cm} \right)
               + {\cal O}(N^{-1}) \,.
\end{eqnarray}

We remark that up to now we have not used the specific form for the
transition probabilities $\omega_{\nu}$. Other expressions for
$\omega_{\nu}$, satisfying detailed balance could also be employed.

Using the partitioning of the network into subsets $I_{\bk}$ we can replace
the sum over $i$ by $\sum_{\bk}\sum_{i\in I_{\bk}}$. Employing the specific
form (\ref{eq:trans}) for $\omega_{\nu}$ and performing the sum over
$\{\bs ,\bsigma\}$ we arrive at (compare \cite{bm})
\begin{equation}
 \frac{\partial \langle \phi\rangle_t}{\partial t}
= \int d\bmu \, \phi(\bmu)
   \sum_{\alpha ,\bk}\frac{\partial}{\partial \mu_{\alpha ,\bk}}
    P(\bmu;t)\left[\frac{1}{3}\mu_{\alpha ,\bk}
             - f_{\alpha ,\bk}(\{\mu_{\beta,\bk}\})\right]\,,
\end{equation}
where $f_{\alpha ,\bk}(\{\mu_{\beta,\bk}\})$ is a function of all the
suboverlaps given by
\begin{eqnarray}
  f_{\alpha ,\bk}(\{\mu_{\beta,\bk}\})
    &&=\frac{1}{12}\{\tanh (\beta L_{\alpha}+\beta L_{\nu})
     +\tanh (\beta L_{\alpha}-\beta L_{\nu})
     +\tanh (\beta L_{\alpha}+\beta L_{\rho})
     \nonumber\\
    && + \tanh (\beta L_{\alpha}-\beta L_{\rho})
     + \mu_{\nu ,\bk}[\tanh (\beta L_{\alpha}+\beta L_{\rho})
     -\tanh (\beta L_{\alpha}-\beta L_{\rho})]
     \nonumber\\
    && + \mu_{\rho ,\bk}[\tanh (\beta L_{\alpha}+\beta L_{\nu})
     -\tanh (\beta L_{\alpha}-\beta L_{\nu})]\}
\end{eqnarray}
with the $L_{\alpha}$ ($\alpha,\nu,\rho =1,2,3$ and different from
each other) given by eq.~(\ref{eq:els}).
Since this equation holds for every smooth function $\phi$ the
equation for the probabilities $P(\bmu;t )$ has the form
\begin{equation}
 \frac{\partial P(\bmu;t )}{\partial t}
 =\sum_{\alpha ,\bk}
    \frac{\partial}{\partial \mu_{\alpha ,\bk}}P(\bmu;t )
    \left[\frac{1}{3}\mu_{\alpha ,\bk}
                  -f_{\alpha ,\bk}(\{\mu_{\beta,\bk}\}) \right]
\end{equation}
and the corresponding flow equations for the submagnetisations
$\mu_{\alpha ,\bk} $ themselves read
\begin{equation}
  \frac{\partial \mu_{\alpha ,\bk}}{\partial t}
    =-\frac{1}{3}\mu_{\alpha , \bk}+f_{\alpha ,\bk}(\{\mu_{\beta,\bk}\})
          \,. \label{eq:muev}
\end{equation}

Together with (\ref{eq:qmu}) these coupled equations are used to study the
dynamic behaviour of the network. In the following section the
results of this study are discussed.

\section{Results}
\label{sec:res}
\pin
In this section we discuss the numerical results for the ATNN model
obtained from the fixed-point equations specifying the thermodynamic
properties and from the flow equations for the suboverlaps describing the
dynamics. We treat the cases of linked patterns and unlinked patterns
separately. We report the results for a set of representative examples
illustrating the main new features of the model.

\subsection{Unlinked patterns}
\begin{figure}[h]
\epsfysize=6cm
\epsfxsize=12cm
\centerline{\epsfbox{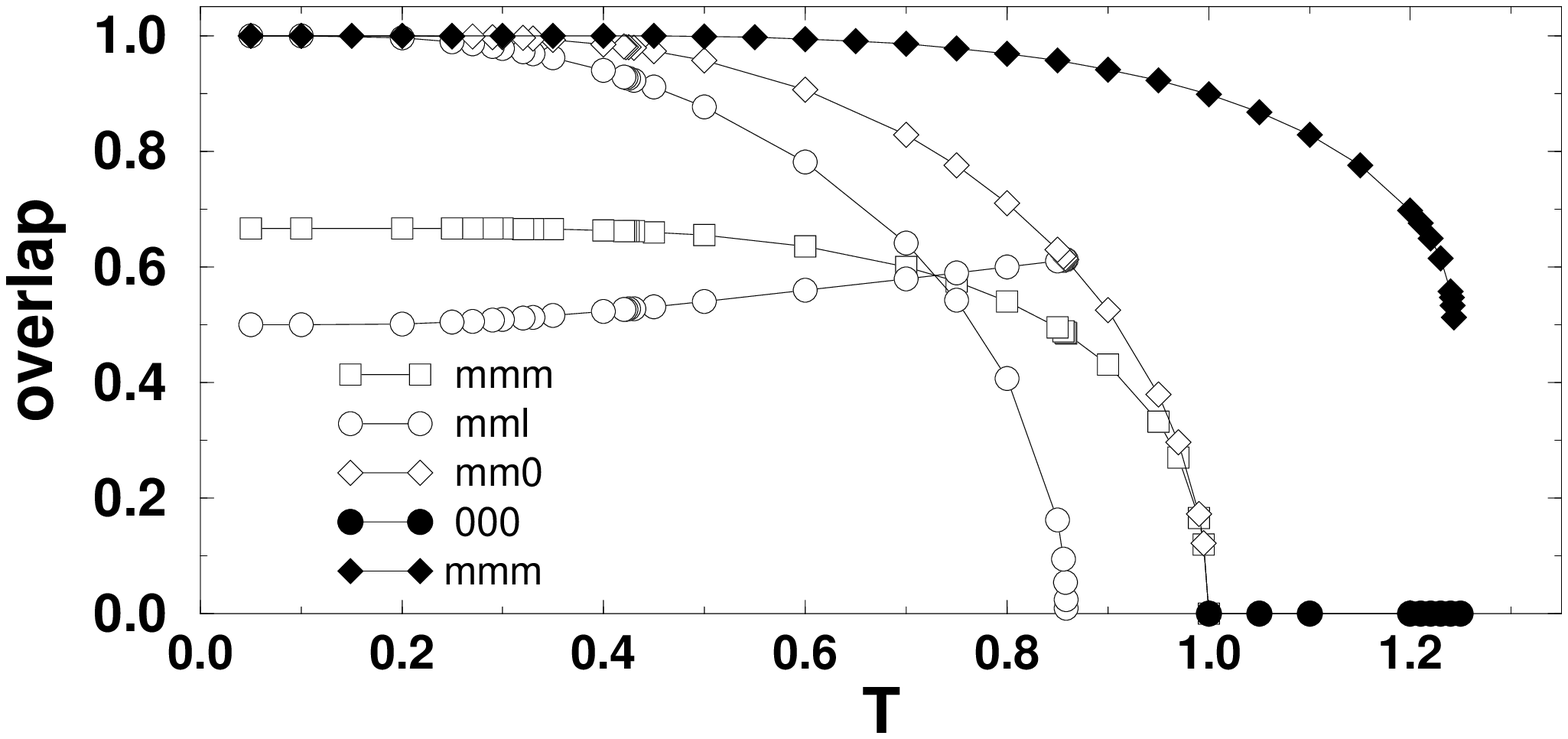}}
\caption{The overlaps for the Mattis states as a function of
the temperature $T=1/\beta$ for $J_1=J_2=J_3=1$. States
represented by filled symbols are only present for the model with linked
patterns.} \label{sol1}
\end{figure}

\pin
We first consider the model with equal coupling parameters
$J_3=J_1(=J_2)=1$.
Introducing Mattis-type states $m_1m_2m_3$
into the fixed-point equations (\ref{eq:ec2}) we obtain the same
equations for $\bm_{\alpha}$, $\alpha =1,2,3$ irrespective of the degree of
dependence between the different types of patterns.
The solutions are presented in Fig.~\ref{sol1} where the overlap for
different Mattis states is shown as a function of the temperature
$T=1/\beta $.
A stability analysis performed by studying the stability matrix given by
$\mbox{A}^{\mu \nu}_{\alpha \beta}= \frac{\partial^2 f}
      {\partial m_{\alpha}^{\mu} \partial m_{\beta}^{\nu}}$
leads to 
the following main features. Above $T=1.0$ there exist no stable
Mattis states. Furthermore, the (paramagnetic) state $000$ where all the
overlaps with the embedded patterns are zero is stable.

Below $T=1.0$ we have the retrieval phase with many different forms of
stable Mattis states. We expect that the most important ones are those
with the lowest energies, i.e. the states $mmm$ (for temperatures in the
interval $(0.83,1)$)
 and $mm0$ (for temperatures in $(0,0.83)$). The state $mm0$
corresponds to a situation where the overlaps with a pattern in the
$s$-part and $\sigma$-part of the network are non-zero and equal to each
other and the overlap with a pattern in the $s\sigma$-part of the network
is zero. This state
is completely equivalent to the states $m0m$ and $0mm$, a fact resulting
directly from the symmetry of the model. Its properties are analogous
to those of the Mattis states in the Hopfield model \cite{ags}, since the
fixed-point equations for the overlap $m$ are the same.
The states which have no analogues in the Hopfield model are $mmm$ and
also $mml$. The first can be interpreted as the description of retrieval
of one pattern, ``simultaneously by the $s, \sigma$ and $s\sigma$-parts of
the network" (for the case of dependent patterns, i.e., case 1 (c), this
pattern is the same for the three parts). However, such a retrieval occurs
with a lot of errors as can be inferred from the small values of the
corresponding overlap in Fig.~\ref{sol1}.
The second state, i.e. $mml$, which is, of course, equivalent to $mlm$ and
$lmm$ differs from the $mmm$ state in this respect that the non-zero
overlaps with the pattern in the three different parts of the network are
not equal to each other. But this state does not seem to play an
important role because it is never a global minimum in the set of
Mattis states (see Fig.~\ref{sol1}).

Increasing the temperature to $T=1$ we notice a continuous transition from
the network retrieval phase to the disordered (paramagnetic) phase.

Next, we have also studied the local minima structure  for non-equal
 values of the coupling parameters $J_1$,
$J_2$ and $J_3$. It only differs in a quantitative way.

\begin{figure}[h]
\epsfysize=5cm
\epsfxsize=10cm
\centerline{\epsfbox{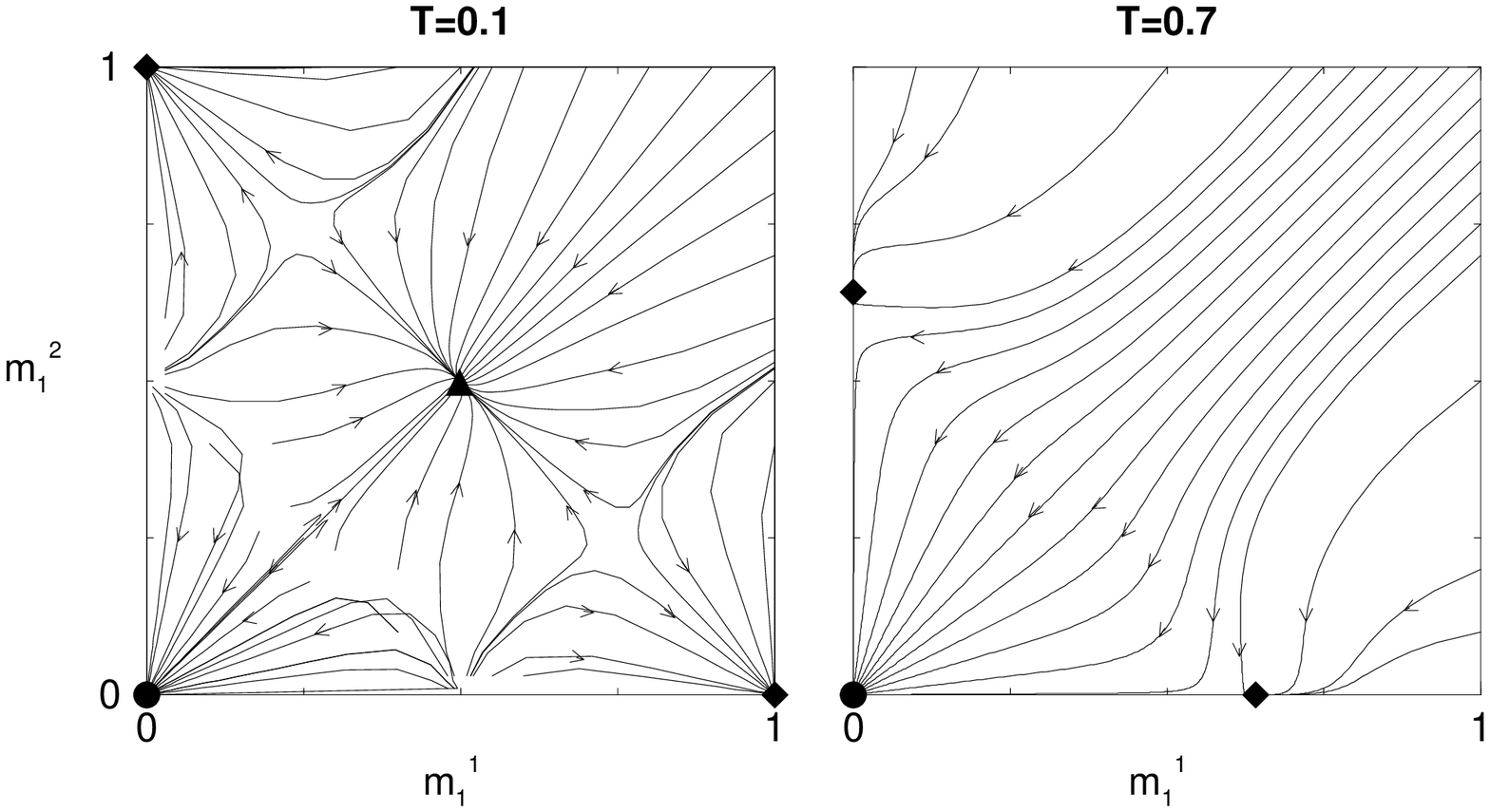}}
\epsfysize=5cm
\epsfxsize=10cm
\centerline{\epsfbox{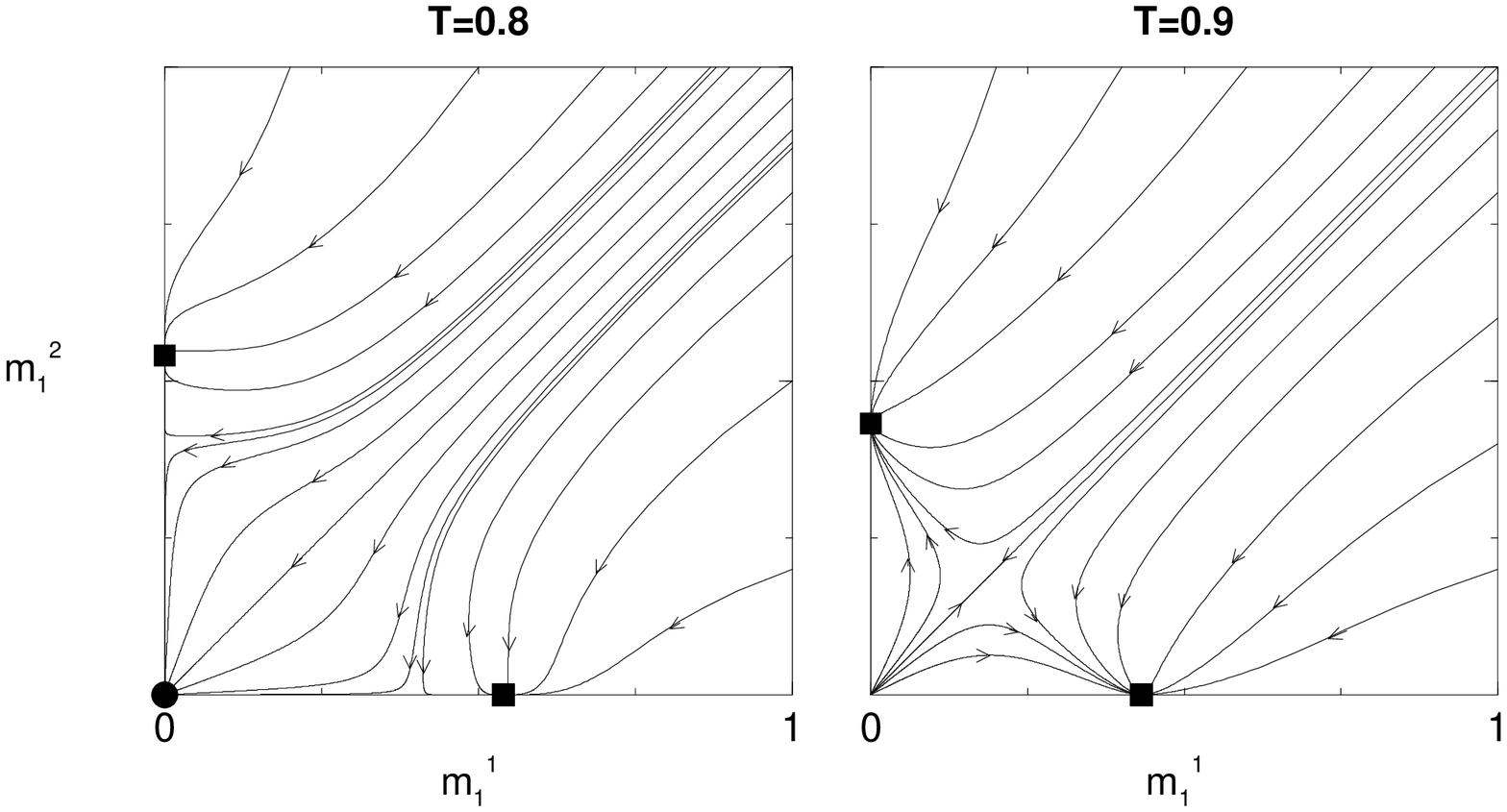}}
\caption{Flow diagrams for the model with unlinked patterns at
several temperatures.
Filled symbols denote different states: a circle for $0mm$, a
diamond for $lmm$, a square for $mmm$ and a triangle for the symmetric
state $smm$ defined in the text.}\label{dynuc1}
\end{figure}

The static results found above are confirmed by a study of the dynamic
behaviour of the ATNN model using the coupled eqs.~(\ref{eq:muev}). For
simplicity, we take the number of embedded patterns of each type $p=2$.
Since we have three different kinds of such patterns the results
concern a six-dimensional flow. Some representative two-dimensional
projections are presented in Fig.~\ref{dynuc1}.

As a starting point we take $\bm_2=\bm_3=(0.5,0)$ and different values for
$\bm_1$. Such a choice of initial conditions is not very specific because
of the symmetry properties of the model. It allows us to show some typical
behaviour of the network. An extensive search confirms that other initial
conditions lead only to quantitatively different diagrams. We remark that
the part of the diagrams not shown explicitly is symmetric with respect to
the $m^1_1$ or $m^2_1$ axis.
We choose some relevant values of $T$ suggested by the thermodynamics.
 We can locate in the first diagram of
Fig.~\ref{dynuc1} ($T=0.1$) the attractor $0mm$ in the lower left corner.
In the lower right (and, since there is symmetry with respect to the
off-diagonal also upper left) corner we see the state $lmm$, which looks
like an attractor.
However, our static analysis reveals that they are only saddle points in
the full six-dimensional space. On the off-diagonal we have a state
of the form $\bm_1=m_1(1,1)$, $\bm_2=m_2(1,0)$ and $\bm_3=m_3(1,0)$
denoted by $smm$, i.e., symmetric with respect to the $s$-part of the
network. We remark
that in this diagram some lines cross each other which is caused by the
fact that only the evolution of two order parameters is shown whereas
the third order parameter, $m_2^1=m_3^1$, is also evolving.
One could easily imagine a three-dimensional picture with $m_2^1=m_3^1$
taken as the third coordinate.

For increasing $T$ the $smm$ state is no longer present and the $lmm$
states move along the $m_1^1$-axis (respectively $m_1^2$-axis) until they
disappear for $T \approx 0.8$.
For these temperatures there is a very small difference in free energy
between the various Mattis-type states. This could be the reason that for
$T=0.8$ we were no longer able to detect the states $lmm$ which should
still
exist according to the static analysis. Indeed, at this temperature the
overlap for the $lmm$ states is almost equal to the one for the $mmm$ state
(see Fig.~\ref{sol1}) showing that they are almost identical. Furthermore,
for $T \approx 0.8$ the states $mmm$ appear, move towards the origin on the
$m_1^1$-axis (respectively the $m_1^2$-axis) as seen on the diagram for
$T=0.9$ and disappear for $T=1.0$. From this temperature onwards only the
origin, i.e. the $000$ state is an attractor.

\subsection{Linked patterns}
\pin
Next, we have analysed the model with linked patterns satisfying
$\gamma^{\mu}=\xi^{\mu}\eta^{\mu}$ with $\bxi$ and $\bneta$
i.i.d.r.v., and with equal coupling parameters $J_3=J_1(=J_2)=1$.

Introducing again Mattis-type states $m_1m_2m_3$ into the fixed-point
equations (\ref{eq:ec2}) and checking their stability we find that there
are only two stable solutions: the retrieval state $mmm$ which is stable
below $T=1.213$ and the paramagnetic $000$ state which is stable
above $T=1.0$. The corresponding retrieval overlap is shown in
Fig.~\ref{sol1} (filled symbols). We notice that in contrast to the
model with unlinked patterns a state of the form $mmm$ has a much
bigger overlap.

Hence, we can distinguish different phases: a retrieval phase below
$T=1.213$ and a paramagnetic phase above $T=1.213$. We remark that the
transition at $T=1.213$ is first order. In the temperature region
$1\leq T < 1.213$ both the paramagnetic and Mattis solutions are local
minima of the free energy. Such a region has also been seen in the Potts
model but not in the Hopfield model \cite{bm}.
Finally, we find that Hopfield-type solutions, i.e., Mattis states of
the form $m00$ are (only) saddle points below $T=1.0$.

A detailed study of the flow equations (\ref{eq:muev}) reveals a much more
complicated local minima structure for this model. It turns out that for
the model with linked patterns we still have to distinguish between the
Mattis solutions according to the relative place of the non-zero overlap
components for the different order parameters. Consequently, we
introduce {\it simple} Mattis states where only the same components of the
different order parameters are non-zero, e.g.,  $\bm_1=m_1(1,0)$,
$\bm_2=m_2(1,0)$, and $\bm_3=m_3(1,0)$ for $p=2$, denoted as before by
$m_1m_2m_3$. For equal components $m_{\alpha }$ these are the states we
have encountered in the thermodynamic analysis. (They are equivalent to
$(0,m)(0,m)(0,m)$ and also to $(-m,0)(-m,0)(m,0)$).

Besides, we define {\it crossed} Mattis states where never the same
components of the different order parameters are non-zero, e.g.,
$(m,0)(0,m)(0,0)$ for $p=2$ and $(m,0,0)(0,m,0)(0,0,m)$ for $p=3$. At this
point we remark that a state of the form $(m,0)(0,m)(m,0)$ is neither
simple nor crossed but of a mixed form. For $p=2$ we did not detect the
latter.
The crossed states are in fact equivalent to Mattis states for a model
with unlinked patterns. This can easily be checked by introducing this type
of solutions in the fixed-point equations for the order parameters and
taking appropriate averages over the linked patterns.

\begin{figure}[h]
\epsfysize=5cm
\epsfxsize=10cm
\centerline{\epsfbox{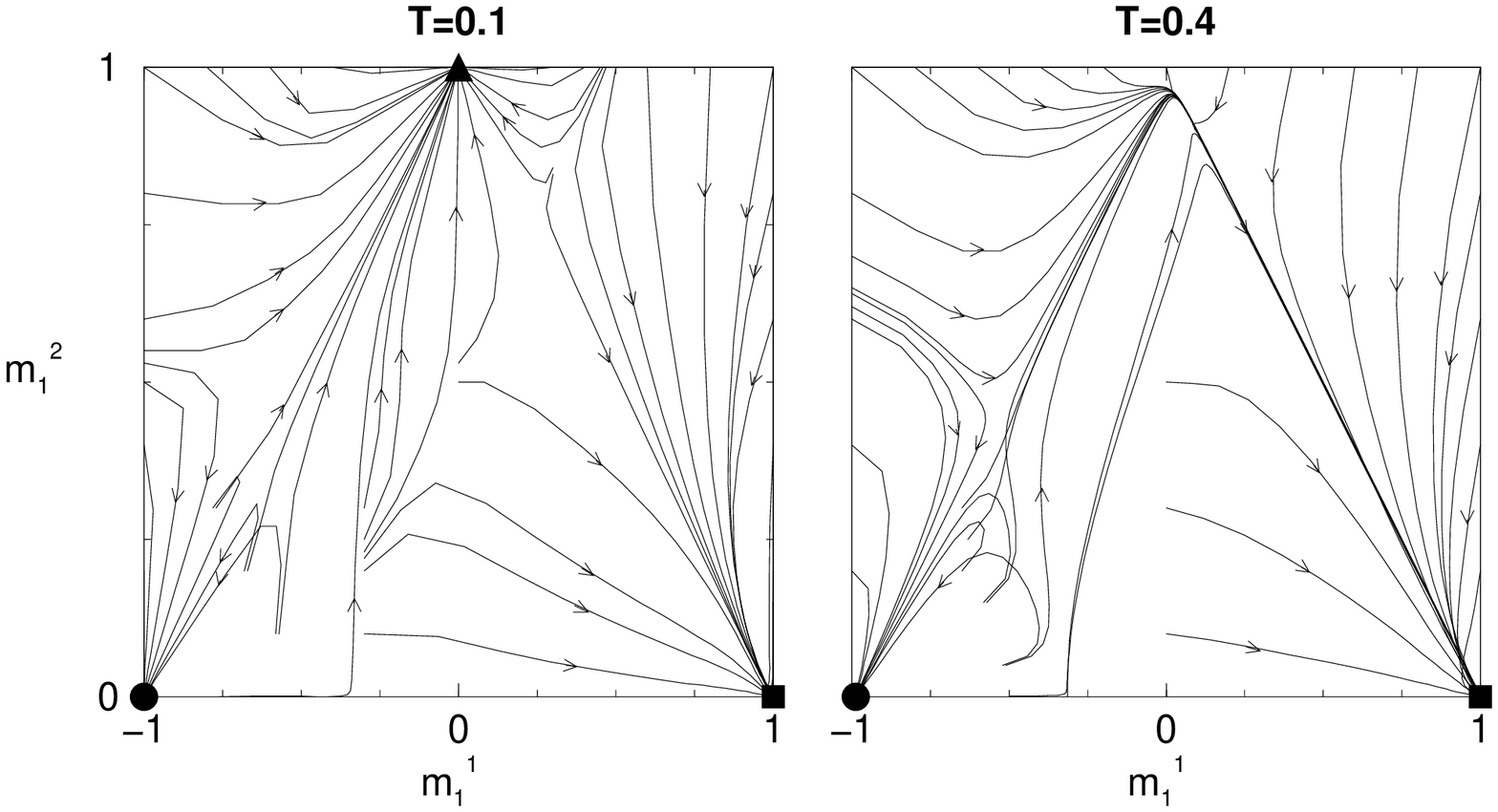}}
\epsfysize=5cm
\epsfxsize=10cm
\centerline{\epsfbox{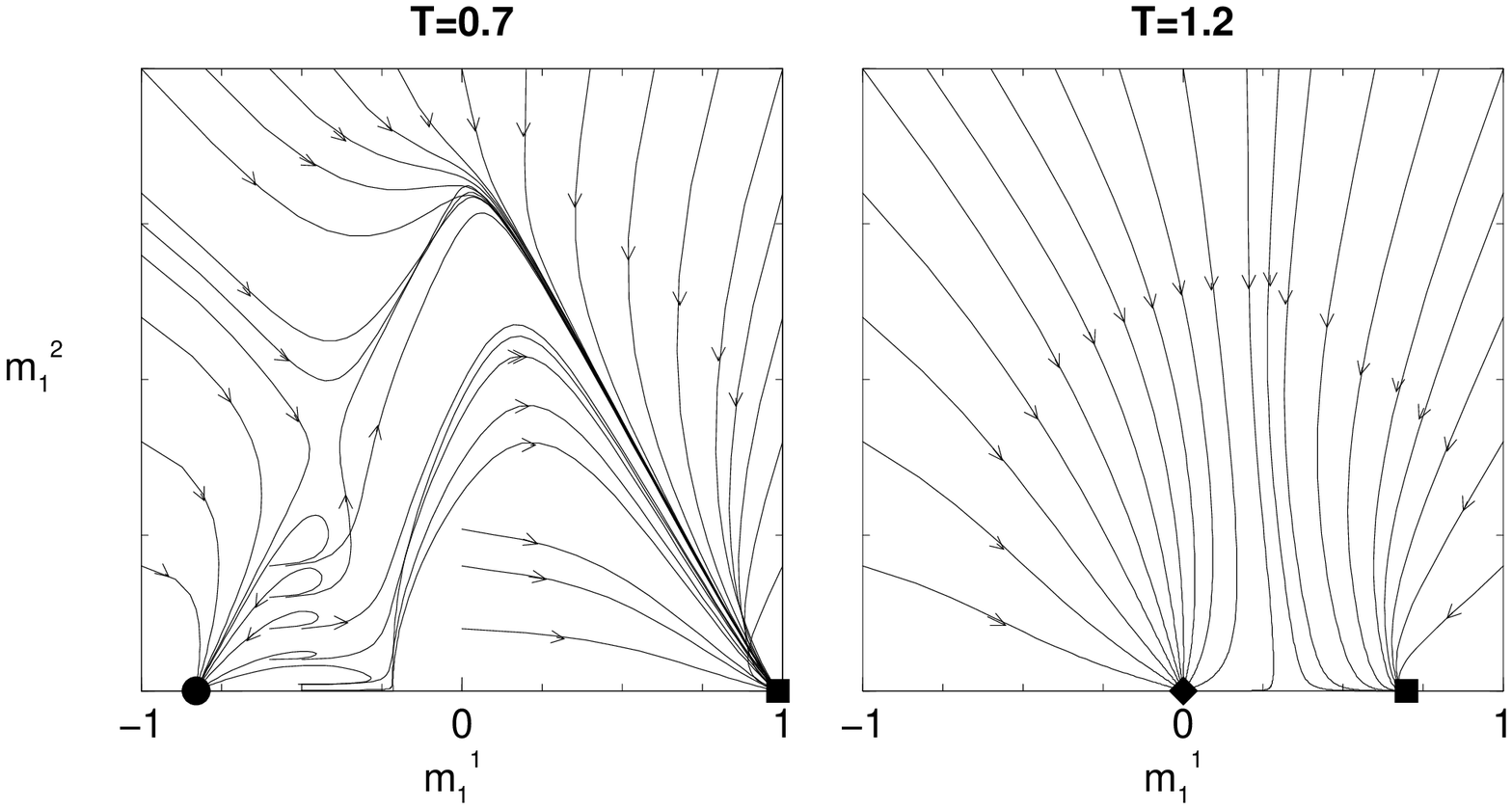}}
\caption{Flow diagrams for the model with linked patterns at different
temperatures. Filled symbols stand
for different states: a circle for $m00$, a square for
$mmm$, a diamond for $000$ and a triangle for the asymmetric state $amm$
defined in the text.}\label{dync1}
\end{figure}

\begin{figure}[h]
\epsfysize=5cm
\epsfxsize=10cm
\centerline{\epsfbox{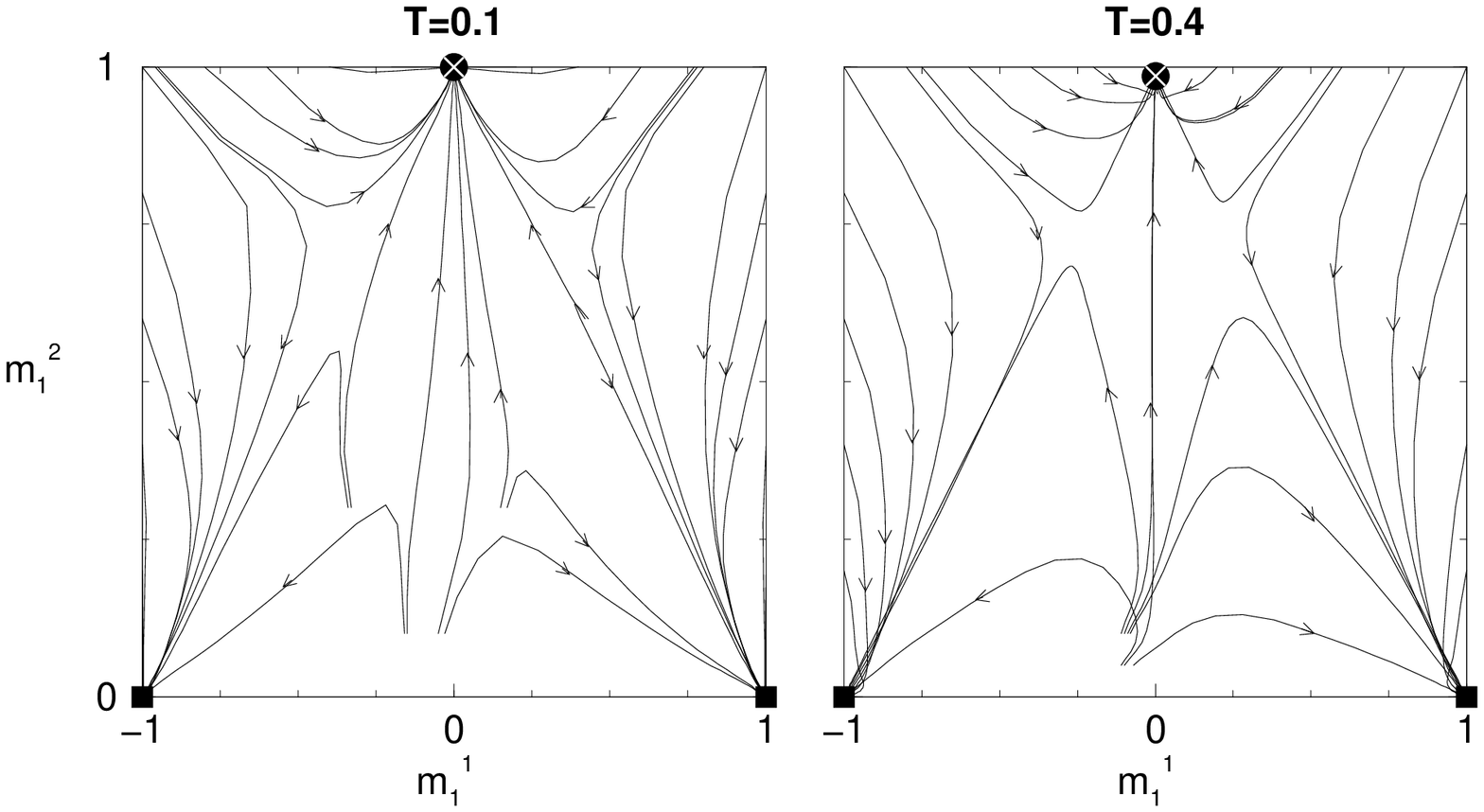}}
\epsfysize=5cm
\epsfxsize=10cm
\centerline{\epsfbox{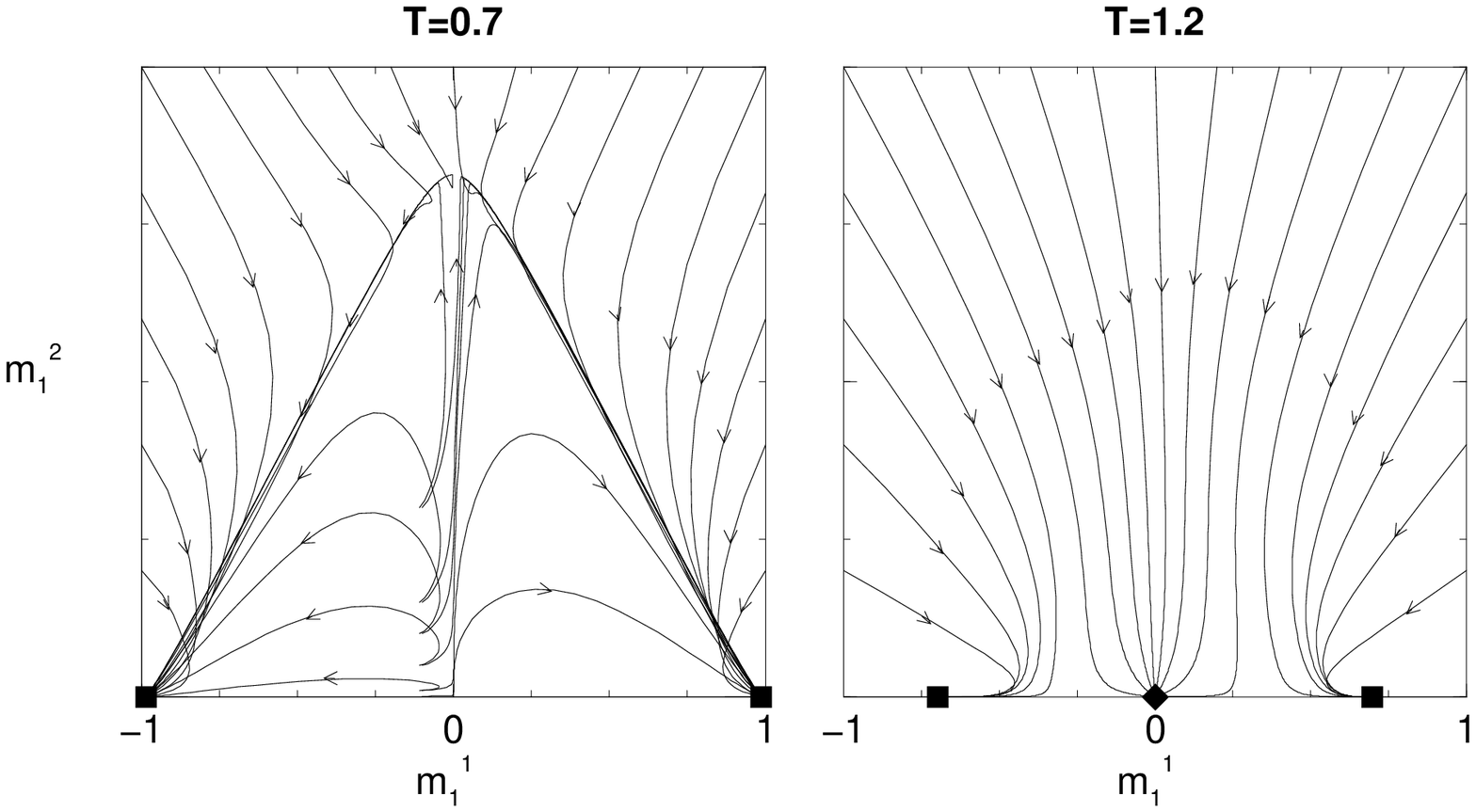}}
\caption{Flow diagrams for the model with linked patterns at different
temperatures. Filled symbols stand for different states: a square for
$mmm$, a diamond for $000$ and a crossed circle for $mm0$.}\label{dync2}
\end{figure}

Some representative flow diagrams are shown in Fig.~\ref{dync1} and
Fig.~\ref{dync2}. As before we present only projections onto a two-dimensional
space. The symmetry of the model with linked patterns is different from the
model with unlinked patterns what results in a different symmetry of the
flow diagrams. So, the remaining part of a diagram in these figures can be
obtained by a reflection of the part displayed with respect to the axis
$m_1^2=0$.

The initial conditions for the flow diagrams of Fig.~\ref{dync1} are as
follows: two identical Mattis states with $\bm_2 =\bm_3=(0.5,0)$.
We can locate in the first diagram of Fig.~\ref{dync1} ($T=0.1$) the
attractor $mmm$ in the lower right corner. In the lower
left corner we see the state $m00$, which again looks like an attractor.
However, it is only a saddle point in the full six-dimensional space.
On the top in the middle we have a state
of the form $\bm_1=(m_1^1,m_1^2)$, $\bm_2=m_2(1,0)$ and $\bm_3=m_3(1,0)$
denoted by $amm$, i.e., asymmetric with respect to the $s$-part of the
network. We remark that also here any crossings of paths are caused by the
fact that the diagrams of Fig.~\ref{dync1} and also of Fig.~\ref{dync2}
are projections of a higher dimensional flow. They are not present in the
full six-dimensional space.

For higher $T$ the state $amm$ disappears, the state $mmm$ stays in the
lower right corner up to $T=1$ and the state $m00$ moves towards the
origin and disappears at $T=1$. Above $T=1$ (see the diagram for
$T=1.2$) both the origin and the state $mmm$ are stable, but the latter
has already moved towards the origin. We note that in contrast with the
model with unlinked patterns, it does not reach
the origin since the transition (to the paramagnetic phase) is first order.

Another illustrative set of diagrams is presented in Fig.~\ref{dync2}. Here
the initial conditions are more general: $\bm_2=(0.1,0)$,
$\bm_3=(0.5,0.0)$.
In the first diagram for $T=0.1$ an attractor $mmm$ is present in the
lower left and right corner. On the top in the middle the state $mm0$ is
located.
It is a crossed state and a minimum at low temperatures. The overlap
$m$ in this state depends on $T$ in the same way as the overlap
of a Mattis solution of the standard Hopfield model does.

For higher $T$ this crossed state disappears but the $mmm$ states
nearly stay at
the same place until $T=1$. Above, the origin is an attractor and the $mmm$
states start to move towards the origin (see the diagram for $T=1.2$). As
explained above they do not reach the origin.

The interesting conclusion of these figures is that the state $mmm$ has a
big basin of attraction. The latter is, of course, somewhat reduced in the
temperature region where also the stable state $000$ appears. Furthermore,
the state $mmm$ has a large overlap (recall Fig.~\ref{sol1}) with the
embedded patterns.

Since the ATNN model with linked patterns seems to have very good
retrieval properties it is worthwhile to derive a $\beta J_1-\beta J_3$
phase diagram.
We first note that the fixed-point equations (\ref{eq:ec2}) for the simple
Mattis states have the same form as those for the mean-field Ashkin-Teller
model. Because these states are always the global minima of the free
energy they determine the transition lines. Of course the meaning of the
phases is different from the standard Ashkin-Teller model. For the special
case of one embedded pattern, i.e., $p=1$, with $\xi_i^1=\eta_i^1=1$ both
models are completely equivalent.

\begin{figure}[h]
\epsfysize=7cm
\epsfxsize=7cm
\centerline{\epsfbox{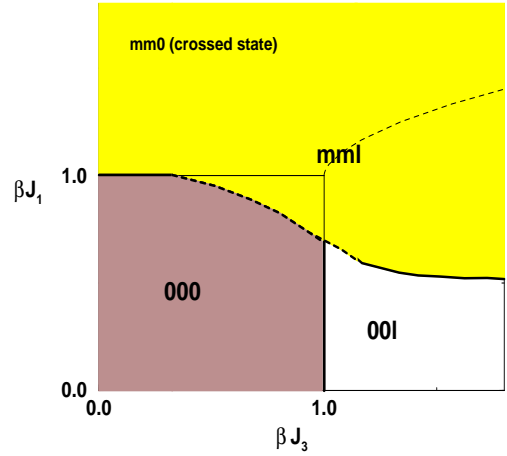}}
\caption{The $\beta J_1-\beta J_3$ phase diagram of the ATNN model with linked
patterns. The dark grey area represents the $000$ phase, the light grey
area the full retrieval phase $mml$ and the white area the partial
retrieval phase $00l$. Thick full lines indicate continuous transitions,
thick dashed lines discontinuous ones.}
\label{missing}
\end{figure}

The ATNN $\beta J_1-\beta J_3$ phase diagram (for low loading of linked
patterns)
is presented in Fig.~\ref{missing}. We distinguish the following phases. The
area in dark grey is
the paramagnetic $000$ phase. The light shadowed area is the full retrieval
phase (for linked patterns) described by the simple Mattis state $mml$
($m \neq l$ when $J_1 \neq J_3$). The white area represents a
partial
retrieval phase, i.e., only patterns embedded in one part of the network
are recalled by the state $00l$. The (thick) full lines indicate
second order transitions, the (thick) dashed lines discontinuous ones. We
remark that the $000$ state exists as a minimum up to the thin full lines,
but outside the dark grey region its energy is higher than the energy of
the $mml$ state. In part of the full retrieval phase, namely in the upper
half of the phase diagram, the crossed state $mm0$ exists. It is stable
in the area above the thin full and thin dashed lines.
\section{Conclusions}
\label{sec:con}
\pin
We have analysed a neural network version of the Ashkin-Teller spin-glass
model for low loading of patterns. Both the thermodynamic and dynamic
properties have been considered, especially for Mattis states which are
the most interesting states from the point of view of retrieval.
Fixed-point equations as well as flow equations for the relevant order
parameters have been derived. Numerical results have been discussed
illustrating the typical behaviour of the network.

The following main conclusions can be drawn. For unlinked embedded patterns
the behaviour of the model is much richer than in the case of the standard
Hopfield model in the sense that many different forms of stable Mattis
states are possible. These states exist up to $T=1$ where a continuous
transition occurs from the retrieval phase to the paramagnetic phase. The
corresponding flow diagrams
are quite complicated but verify the existence of these many attractors.
However, none of these retrieval states has a bigger overlap than the
Mattis states of a corresponding Hopfield model.
Hence the inclusion of the four-neuron interaction term does not
particularly improve the quality of retrieval.

\begin{figure}[h]
\epsfysize=4cm
\epsfxsize=8cm
\centerline{\epsfbox{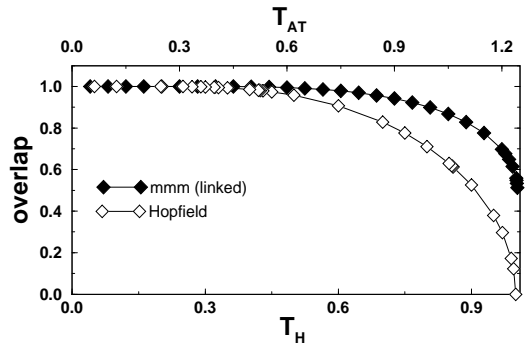}}
\caption{The overlap as a function of a rescaled $T$ for the state $mmm$ in
the ATNN model with linked patterns (filled diamond) versus the
overlap for the Mattis state in the Hopfield model (empty diamond).
Respective temperatures are denoted by $T_{AT}$ and $T_H$.}
\label{res}
\end{figure}

For linked embedded patters interesting new features show up.
The most important one is that stable Mattis states of the form $mmm$
appear. They have a very big overlap with the embedded patterns, meaning
that the pairs of patterns which are linked by the four-neuron term are
retrieved with a very high accuracy. Furthermore, they exist up to
$T=1.213$ and have a big basin of attraction. For temperatures $1\leq
T\leq 1.213$ both these Mattis states and the paramagnetic
solution are local minima of the free energy such that their basin of
attraction is somewhat reduced. To verify then that this big overlap $m$
is not just a rescaled overlap of the corresponding Mattis state of a
corresponding Hopfield model we have made a comparison in Fig.~\ref{res}. We
clearly see the difference in shape in favour of the ATNN.
Further details of additional features of the ATNN model are given in a
$\beta J_1-\beta J_3$ phase diagram (see Fig.~\ref{missing}).
In brief, the linked pairs are retrieved easily and with a high precision
 (simple
states) and the unlinked pairs may be retrieved (crossed states), but
always with lower precision than that of the linked ones.
It is important to stress that the patterns which are linked can be
completely different. In principle, they are independent.

\section*{Acknowledgements}
\pin
This work has been supported in part by the Research Fund of the
K.U.Leuven (Grant OT/94/9).
The authors are indebted to Marc Van Hulle of the Neurophysiology
Department of the K.U.Leuven for interesting discussions concerning the
possible biological relevance of this model. One of us (P.K) would like
to thank Prof.~G. Kamieniarz for encouragement to study neural
networks.  
Both authors acknowledge the Fund for Scientific Research-Flanders
(Belgium) for financial support.

\end{document}